\documentclass[runningheads]{svmult}

\usepackage{graphicx}  % standard LaTeX graphics tool
\usepackage{physprbb}  % modified textarea for proceedings,
\usepackage{verbatim}

\begin{document}
\title*{Precision Spectroscopy of Molecular Hydrogen Ions:
Towards Frequency Metrology of Particle Masses\footnote{This text
will appear in "Precision Physics of Simple Atomic Systems",
{\em Lecture Notes in Physics}, Springer, 2007}}
\toctitle{Precision Spectroscopy of Molecular Hydrogen Ions:
\newline toward Frequency Metrology of Particle Masses}
% allows explicit linebreak for the table of content
%
%
\titlerunning{Spectroscopy of the Hydrogen Molecular Ions.}
% allows abbreviation of title, if the full title is too long
% to fit in the running head
%
\author{
Bernhard Roth\inst{1}
\and Jeroen Koelemeij\inst{1}
\and Stephan Schiller\inst{1}
\and Laurent Hilico\inst{2,3}
\and Jean-Philippe Karr\inst{2,3}
\and Vladimir Korobov\inst{4}
\and Dimitar Bakalov\inst{5}
}
\authorrunning{Bernhard Roth et al.}
% if there are more than two authors,
% please abbreviate author list for running head
%
%
\institute{
Institut f\"ur Experimentalphysik,
  Heinrich-Heine-Universit\"at D\"usseldorf\\
  40225 D\"usseldorf, Germany\\
\and D\'epartement de Physique et Mod\'elisation,
  Universit\'e d'Evry Val d'Essonne\\
  Boulevard F. Mitterrand, 91025 Evry cedex\\
\and Laboratoire Kastler Brossel, Universit\'e Pierre et Marie Curie\\
  T12, Case 74, 4 place Jussieu, 75252 Paris, France
\and Joint Institute for Nuclear Research, 141980 Dubna, Russia
\and Institute for Nuclear Research and Nuclear Energy,
  Sofia 1784, Bulgaria
}

%\label{21}

\maketitle              % typesets the title of the contribution

\begin{abstract}
We describe the current status of high-precision ab initio calculations of
the spectra of molecular hydrogen ions (H$_2^+$ and HD$^+$) and of two
experiments for vibrational spectroscopy. The perspectives for a
comparison between theory and experiment at a level of 1 ppb are
considered.
\index{Hydrogen molecular ions}
\end{abstract}

\section{Introduction}
The molecular hydrogen ion (MHI) is the simplest stable molecule,
containing just two nuclei and a single electron. Since the birth of the
field of molecular physics it has played an important role: it is on one
hand an important benchmark system for detailed studies of energy levels
\cite{22Moss}, for collisions and chemical reactions between charged
molecules and neutral atoms/molecules, of interactions with laser
radiation and energetic charged particles, and for testing the respective
theoretical descriptions. On the other hand, the MHI is also an
astrophysically important molecule, involved in reaction chains leading to
the production of polyatomic molecules. Over 800 publications have been
written on this molecule in the last 35 years \cite{22diref}. The large
majority are theoretical studies.

Concerning high-resolution spectroscopy of MHIs, only a limited number of
investigations have been carried out, most of which a long time ago.
Radiofrequency spectroscopy of the hyperfine structure in several
vibrational levels has been performed on H$_2^+$ trapped in a Paul trap
\cite{22Jefferts}; several low-lying fundamental ro-vibrational transitions
of HD$^+$ have been measured using laser spectroscopy on an ion beam
\cite{22Wing}, while rotational and ro-vibrational transitions of H$_2^+$,
D$_2^+$ and HD$^+$ close to the dissociation limit were investigated using
microwave and laser spectroscopy, also on an ion beam
\cite{22Carrington,22Critchley}.  The highest spectroscopic accuracies
reported so far were achieved in the experiments of Jefferts and of Wing
{\em et al.} \cite{22Wing,22Spezeski}, $\simeq\!1\!\cdot\!10^{-6}$ in relative
units. Recently, also the dissociation energies have been obtained with
accuracies $\simeq1\cdot10^{-6}$ \cite{22Eyler}. Thus, the experimental
accuracies have been far less than those achieved in hydrogen or helium
spectroscopy.

\index{Hydrogen molecular ions!ro-vibrational spectroscopy}
\index{Doppler-free spectroscopy}
In the late 1990s, it was recognized that there are attractive reasons and
many opportunities to study MHIs in novel ways and to achieve a much
higher precision than previously possible \cite{22recognition}. Several
techniques, not used before on MHIs, appeared to be applicable, including
translational cooling, internal cooling, spectroscopy with reduced Doppler
broadening, Doppler-free spectroscopy, high sensitivity ion detection.
Novel laser systems not available at the time of the last precision
spectroscopic studies can be used advantageously, among them diode lasers,
quantum cascade lasers and the femtosecond frequency comb. The prospect of
significantly improved experimental precision has also motivated us to
develop more extended theoretical treatments of the MHI; in the course of
these efforts, the accuracy of the energy levels has been increased by
approximately two orders of magnitude compared to previous work.

Some of the above techniques have by now been implemented, and are
reported here; the remaining appear to be feasible in the near
future. These recent developments open up a number of novel
applications of MHIs:

\begin{description}
\item[(i)] test advanced {\em ab initio} molecular calculations
  (in particular, QED contributions)

\item[(ii)] measure certain fundamental constants

\item[(iii)] test concepts for the manipulation of molecules (state
preparation, alignment)

\item[(iv)] sense fields (blackbody radiation \cite{22Koelemeij06})

\item[(v)] probe fundamental physics laws (e.g. Lorentz Invariance
\cite{22Mueller04}, time invariance of fundamental constants
\cite{22Schiller05,22review})
%\index{Variation~of~fundamental constants!memp@$m_e/m_p$}
%\index{Variation~of~fundamental constants!mpmd@$m_p/m_d$}

\item[(vi)] study electric dipole interactions between molecules
\cite{22DeMille}

\item[(vii)] explore elastic, reactive and charge exchange collisions with
neutral atoms and molecules at ultralow collision energies
\end{description}

The successful demonstration of manipulation of MHIs at the quantum state
level could also open up the possibility to study collisions with
quantum-state resolution, i.e. where all parent particles are in specific
quantum states.

\index{Determination of fundamental constants via rovibrational spectroscopy of MHIs!memp@$m_e/m_p$}
\index{Determination of fundamental constants via rovibrational spectroscopy of MHIs!mpmd@$m_p/m_d$}
An attractive perspective  of our work pursued under (i) is to
eventually determine the ratios of electron-to-proton mass ($m_e/m_p$),
proton-to-deuteron mass ($m_p/m_d$) and proton-to-triton mass ($m_p/m_t$)
from a comparison between accurate experimental and theoretical energy
level data. The basis for this possibility is the dependence of the
vibrational and rotational transition frequencies on the fundamental
constants. For fundamental vibrational and rotational transitions, the
frequencies scale approximately as
\begin{equation}
h\,\nu_{vib} \sim \sqrt{m_e/\mu}\>R_\infty\,,\qquad
h\,\nu_{rot} \sim (m_e/\mu)R_\infty\,,
\end{equation}
where $\mu=M_1M_2/(M_1\!+\!M_2)$ is the reduced mass of the two nuclei and
$R_\infty$ is the Rydberg energy. The precise dependencies have been
computed in refs. \cite{22Schiller05,22paris1,22Karr06}. The mass ratios
$m_p/m_d$, $m_p/m_t$ and $m_e/m_p$ are conventionally determined by
Penning ion trap mass spectrometry on single particles or by electron spin
resonance of single hydrogen-like ions in a Penning ion trap. Relative
accuracies are currently $2.0\cdot10^{-10}$ \cite{22CODATA},
$2\cdot10^{-9}$, and $4.6\cdot10^{-10}$ \cite{22CODATA,22Verdu},
respectively. Note that in case of $m_e/m_p$, the determination involves
the use of QED \cite{22Karsh05}. Clearly, the corresponding accuracies of $\nu_{vib}$,
$\nu_{rot}$ represent the goal levels for our ongoing experimental and
theoretical efforts on H$_2^+$ and HD$^+$.

\index{Systematic shifts}
Several aspects support the expectation that such accuracies can be
reached in the near future. First, the lifetimes of vibrational levels are
long, the shortest ones occurring for low-lying levels in HD$^+$,
$\simeq10\,$ms. The relative linewidth due to spontaneous decay is thus of
the order or smaller than $10^{-13}$. Second, Doppler broadening can be
strongly reduced or eliminated by either cooling the molecular ions or by
performing two-photon Doppler-free spectroscopy. Finally, collision
broadening and time-of-flight broadening can also be minimized by both
cooling and providing a good ultra-high vacuum environment. Systematic
shifts due to light fields, trap electric fields, and trap and
environmental magnetic fields will need to be considered; hereby it will
be helpful that these influences be calculated accurately, using the
relative simplicity of the MHI. The theoretical determination of the
energy levels at the goal accuracy level will need as input nuclear
properties such as the proton and deuteron nuclear radii, which may be
obtained e.g. from hydrogen spectroscopy or nuclear scattering
experiments.

In this contribution we present an overview of our theoretical and
experimental results achieved recently on MHIs. Chapter 2 describes the
theoretical approaches for a precise computation of energy levels,
including hyperfine and QED effects and the computation of one- and
two-photon spectra. Chapter 3 presents the development and results from an
experiment for trapping and spectroscopy of H$_2^+$ performed at the
Universit\'e d'Evry Val d'Essonne. Chapter 4 summarizes an experiment
on HD$^+$ at the University of D\"usseldorf.

\section{{\em Ab initio} theory of H$_2^+$ and HD$^+$}

\index{Hydrogen molecular ions!theory}

The dissociation energies of 462 states in H$_2^+$ and 619 in HD$^+$ in a
wide range of $v$ and $L$, vibrational and rotational quantum numbers,
have been calculated some time ago by R.E.~Moss \cite{22moss-H2+,22Moss1993}
with a relative accuracy of $\sim\!5\!\cdot\!10^{-9}$ (including the
leading order relativistic and radiative corrections). Later the numerical
precision of the nonrelativistic energies have been improved up to
$10^{-15}-10^{-24}$ {a.u.}
\cite{22Schiller05,22paris1,22paris2,22Kor00,22Bai02,22Yan03,22Drake04} by using
variational methods. The ultimate accuracy of $\sim\!10^{-24}$ a.u.\ has
been obtained for the $\mbox{H}_2^+$ ground state \cite{22Drake04}. These
calculations demonstrate that at least the nonrelativistic ro-vibrational
transition frequencies can be determined with an uncertainty well below
the 1 kHz level. In this section we describe the calculation of QED
corrections as an expansion in terms of $\alpha$, the fine structure
constant. The numerical method exploits a variational approach based on
the Slater-type exponents as basis functions. We demonstrate that the
frequencies of ro-vibrational transitions can be obtained in this way with
a precision better than 1 part per billion (ppb).

\subsection{Variational expansion}

The variational bound state wave functions are calculated by solving the
three-body Schr\"{o}dinger equation with Coulomb interaction using the
variational approach based on the exponential expansion with randomly
chosen exponents. Details and the particular strategy of choice
of the variational nonlinear parameters and basis structure that have been
adopted in the present work can be found in \cite{22Kor00}.

Briefly, the wave function for a state with a total orbital angular
momentum $L$ and of a total spatial parity $\pi=(-1)^L$ is expanded as
follows:
\begin{equation}\label{22exp_main}
\begin{array}{@{}l}
\displaystyle \Psi_{LM}^\pi(\mathbf{R},\mathbf{r}_1) =
       \sum_{l_1+l_2=L}
         \mathcal{Y}^{l_1l_2}_{LM}(\hat{\mathbf{R}},\hat{\mathbf{r}}_1)
         G^{L\pi}_{l_1l_2}(R,r_1,r_2),
\\[3mm]\displaystyle
G_{l_1l_2}^{L\pi}(R,r_1,r_2) =
    \sum_{n=1}^N \Big\{C_n\,\mbox{Re}
          \bigl[e^{-\alpha_n R-\beta_n r_1-\gamma_n r_2}\bigr]
\\[1mm]\displaystyle\hspace{33mm}
+D_n\,\mbox{Im} \bigl[e^{-\alpha_n R-\beta_n r_1-\gamma_n r_2}\bigr] \Big\}.
\end{array}
\end{equation}
Here $\mathcal{Y}^{l_1l_2}_{LM}(\hat{\mathbf{R}},\hat{\mathbf{r}}_1)=
R^{l_1}r_1^{l_2}\{Y_{l_1}\otimes Y_{l_2}\}_{LM}$ are the solid bipolar
harmonics, $\mathbf{R}$ is the position vector of nucleus 2 relative to
nucleus 1, and $\mathbf{r}_1$, $\mathbf{r}_2$ are positions of an electron
relative to nuclei 1 and 2, respectively. The complex exponents, $\alpha$,
$\beta$, $\gamma$, are generated in a pseudorandom way.

When the exponents $\alpha_n$, $\beta_n$, and $\gamma_n$ are real, the
method reveals slow convergence for molecular type Coulomb systems. The
use of complex exponents allows to reproduce the oscillatory behaviour of
the vibrational part of the wave function and to improve convergence
\cite{22Frolov95,22Kor00}.

The advantage of choice (\ref{22exp_main}) is the simplicity of the basis
functions. It allows evaluating analytically matrix elements of the
Breit-Pauli Hamiltonian and the leading-order radiative corrections and,
more importantly, to treat in a systematic way the singular integrations
encountered in higher-order contributions \cite{22integrals}.

\subsection{Leading-order relativistic and radiative corrections}

Relativistic corrections of the leading $R_\infty\alpha^2$ order, the
Breit-Pauli Hamiltonian, are well known and can be found in many textbooks
\cite{22BS,22BLP}. The nuclear finite size effects are considered as
contributions to this order. Details, relevant particularly to the case of
the MHIs, can be found in \cite{22Kor06rel}.

In what follows we assume that the nuclear charges are
$Z_1\!=\!Z_2\!=\!Z\!=\!1$ and nuclear masses are denoted by capital $M$.
The units adopted are ($\hbar\!=\!e\!=\!m_e\!=\!1$).

The radiative corrections of an order $R_\infty\alpha^3$ for a one
electron molecular system can be expressed by the following set of
equations (see Refs.~\cite{22Pac98,22Yel01,22HD_BL}).

The one-loop self-energy correction (orders $R_\infty\alpha^3$ and
$R_\infty\alpha^3(m/M)$) is:
\begin{equation}\label{22se3}
\begin{array}{@{}l}
\displaystyle
E_{se}^{(3)} =
\frac{4\alpha^3Z}{3}
         \left(
            \ln\frac{1}{\alpha^2}-\beta(L,v)+\frac{5}{6}-\frac{3}{8}
         \right)
         \left\langle
            \delta(\mathbf{r}_1)\!+\!\delta(\mathbf{r}_2)
         \right\rangle,
\\[4mm]\displaystyle\hspace{10mm}
+\alpha^3Z^2\sum_{i=1,2}
    \left[
    \frac{2}{3M_i}
        \left( \!-\ln\alpha\!-\!4\,\beta(L,v)\!+\!\frac{31}{3} \right)
        \left\langle \delta(\mathbf{r}_i) \right\rangle
    -\frac{14}{3M_i}Q(r_i)
    \right],
\end{array}
\end{equation}
where
\begin{equation}\label{22Bethe}
\beta(L,v) =
   \frac{
   \left\langle
      \mathbf{J}(H_0\!-\!E_0)\ln\left((H_0\!-\!E_0)/R_\infty\right)\mathbf{J}
   \right\rangle}
   {\left\langle
      [\mathbf{J},[H_0,\,\mathbf{J}]]/2
   \right\rangle}
\end{equation}
is the Bethe logarithm. The latter quantity presents the most difficult
numerical problem in computation of QED corrections for the three-body
bound states. In \cite{22HD_BL,22H2_BL} the calculations for a wide range of
ro-vibrational states in $\mbox{H}_2^+$ and $\mbox{HD}^+$ have been
performed to an accuracy of about 7 significant digits. The operator
$\mathbf{J}$ in (\ref{22Bethe}) is the electric current density operator of
the system\footnote{$\mathbf{J}\!=\!\sum_a z_a\mathbf{p}_a/m_a$, where
$z_a$, $\mathbf{p}_a$, $m_a$ are the charge, impulse, and mass of a
particle $a$. The sum is performed over all particles of the system.}. The
last term, $Q(r)$, in Eq.~(\ref{22se3}) is the mean value of a regularized
operator introduced by Araki and Sucher \cite{22as} for the $1/(4\pi r^3)$
potential:
\begin{equation}\label{22Qterm}
Q(r) = \lim_{\rho \to 0} \left\langle
            \frac{\Theta(r - \rho)}{ 4\pi r^3 }
      + (\ln \rho + \gamma_E)\delta(\mathbf{r}) \right\rangle.
\end{equation}
The values of this matrix element for ro-vibrational states are calculated
in \cite{22Kor06rel}.

The remaining contributions in this order can be obtained from the Pauli
form factor of an electron (anomalous magnetic moment):
\begin{equation}
E_{anom}^{(3)} = \pi\alpha^2Z
         \left[\frac{1}{2}\left(\frac{\alpha}{\pi}\right)\right]
         \left\langle
            \delta(\mathbf{r}_1)\!+\!\delta(\mathbf{r}_2)
         \right\rangle.
\end{equation}
and from the one-loop vacuum polarization:
\begin{equation}
E_{vp}^{(3)} = \frac{4\alpha^3Z}{3}
         \left[-\frac{1}{5}\right]
         \left\langle
            \delta(\mathbf{r}_1)\!+\!\delta(\mathbf{r}_2)
         \right\rangle.
\end{equation}

\subsection{$R_\infty\alpha^4$ order corrections in the nonrecoil limit}

The contribution of recoil corrections, proportional to $(m/M)$, in the
$R_\infty\alpha^4$ order are too small for our present consideration and
may be neglected. Radiative corrections for a bound electron in an
external field are known analytically \cite{22SapYen,22Eides01}:
\begin{equation}\label{22fourth}
\begin{array}{@{}l}
\displaystyle
E_{se}^{(4)} =
\alpha^4Z^2
      \left[
         4\pi\left(\frac{139}{128}-\frac{1}{2}\ln{2}\right)
      \right]
         \left\langle
            \delta(\mathbf{r}_1)\!+\!\delta(\mathbf{r}_2)
         \right\rangle,
\\[4mm]\displaystyle
E_{vp}^{(4)} = \alpha^4Z^2
         \left[\frac{5\pi}{48}\right]
         \left\langle
            \delta(\mathbf{r}_1)\!+\!\delta(\mathbf{r}_2)
         \right\rangle,
\\[4mm]\displaystyle
E_{anom}^{(4)} = \alpha^2Z\pi
         \left[
            \left(\frac{\alpha}{\pi}\right)^2
            \left(
               \frac{197}{144}+\frac{\pi^2}{12}-\frac{\pi^2}{2}\ln{2}
                  +\frac{3}{4}\zeta(3)
            \right)
         \right]
         \left\langle
            \delta(\mathbf{r}_1)\!+\!\delta(\mathbf{r}_2)
         \right\rangle,
\\[4mm]\displaystyle
E_{2loop}^{(4)} = \alpha^2Z\pi
         \left[
            \left(\frac{\alpha}{\pi}\right)^2
            \left(
               \!-\frac{6131}{1296}-\frac{49\pi^2}{108}+2\pi^2\ln{2}-3\zeta(3)
            \right)
         \right]
         \left\langle
            \delta(\mathbf{r}_1)\!+\!\delta(\mathbf{r}_2)
         \right\rangle.\hspace{-5mm}
\end{array}
\end{equation}
The last equation includes both the Dirac form factor and polarization
operator contributions.

\begin{figure}[t]
\begin{center}
\includegraphics[width=.5\textwidth]{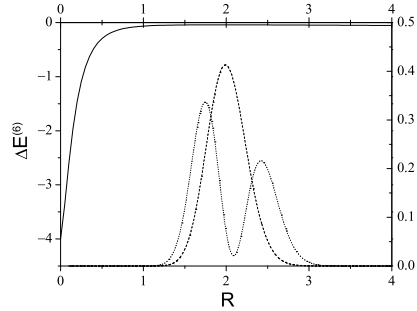}
\end{center}
\vspace{-5mm}
\caption[]{Adiabatic potential of the $m\alpha^6$ order contribution to
the Dirac energy of the two-center problem ($Z_1\!=\!Z_2\!=\!1$). Dashed
curves are the squared densities of the ground and first vibrational state
wave functions of $\mbox{H}_2^+$ ion.} \label{22a6rel}
\end{figure}

The $R_\infty\alpha^4$ relativistic correction is obtained using the
adiabatic "effective" potential for an $m\alpha^6$ term in the $\alpha$
expansion of the two-center Dirac energy (see Fig.~\ref{22a6rel}).
Averaging over the squared wave function density of a state one gets
$E_{rc}^{(4)}$. The adiabatic potentials have been obtained recently with
about 5 significant digits \cite{22Kor_a6}, and the Born-Oppenheimer
solution yields approximate wave functions at the
$(m/M)\!\approx\!10^{-4}$ level, which allows to claim that $E_{rc}^{(4)}$
is now known to 4 digits.

Some higher order radiative corrections for a bound electron in an
external field are also known in an analytic form \cite{22SapYen,22Eides01}
and can be included into consideration:
\begin{equation}\label{22a5}
E_{se}^{(5)} =
\alpha^5Z^3\,\ln^2{\!(Z\alpha)^{-2}}\bigl[-1\bigr]\>
         \left\langle
            \delta(\mathbf{r}_1)\!+\!\delta(\mathbf{r}_2)
         \right\rangle.
\end{equation}

\index{H2plus@H$_2^+$ ions}
\index{HDplus@HD$^+$ ions}
\begin{table}[b]
\caption{Summary of contributions to the
$(v\!=\!0,L\!=\!0)\!\to\!(v'\!=\!1,L'\!=\!0)$
transition frequency (in MHz).} \label{22summary}
\begin{center}
\begin{tabular}{l@{\hspace{5mm}}r@{}l@{\hspace{6mm}}r@{}l}
\hline\hline
\vrule height 10.5pt width 0pt depth 3.5pt
 & \multicolumn{2}{c}{$\mbox{H}_2^+$} & \multicolumn{2}{c}{$\mbox{HD}^+$} \\
\hline
\vrule height 10pt width 0pt% depth 3.5pt
$\Delta E_{nr}$ & 65\,687\,511&.0686    & 57\,349\,439&.9717    \\
$\Delta E_{\alpha^2}$ &   1091&.041(03) &          958&.152(03) \\
$\Delta E_{\alpha^3}$ & $-$276&.544(02) &       $-$242&.118(02) \\
$\Delta E_{\alpha^4}$ &   $-$1&.997     &         $-$1&.748     \\
$\Delta E_{\alpha^5}$ &      0&.120(23) &            0&.106(19) \\
\hline\\[-3mm]
$\Delta E_{tot}$& 65\,688\,323&.688(25) & 57\,350\,154&.368(21)\\
\hline\hline
\end{tabular}
\end{center}
\end{table}

The electron ground state wave function may be approximated by
$\psi_e(\mathbf{r}_e) =
C[\psi_{1s}(\mathbf{r}_1)+\psi_{1s}(\mathbf{r}_2)]$, where $\psi_{1s}$ is
the hydrogen ground state wave function and $C$ is a normalization
coefficient. Thus, one may use this approximation to evaluate other
contributions in the $R_\infty\alpha^5$ order:
\begin{equation}\label{22high}
\begin{array}{@{}l}
\displaystyle
E_{se}^{(5')} =
\alpha^5Z^3
   \Bigl[
       A_{61}\ln{(Z\alpha)^{-2}}
      +A_{60}
   \Bigr]
   \left\langle
      \delta(\mathbf{r}_1)\!+\!\delta(\mathbf{r}_2)
   \right\rangle,
\\[3mm]\displaystyle
E_{2loop}^{(5)} = \frac{\alpha^5}{\pi}Z^2
   \bigl[
      B_{50}
   \bigr]
   \left\langle
      \delta(\mathbf{r}_1)\!+\!\delta(\mathbf{r}_2)
   \right\rangle,
\end{array}
\end{equation}
where the constants $A_{61}$, $A_{60}$, and $B_{50}$ are taken equal to
the constants of the $1s$ state of the hydrogen atom $A_{61}=5.419\dots$
\cite{22Layzer60}, $A_{60}=-30.924\dots$ \cite{22Pac93}, and
$B_{50}=-21.556\dots$ \cite{22b50} (see also Ref.~\cite{22Eides01} and
references therein). The final theoretical uncertainty in the transition
frequency (see Table \ref{22summary}) is determined by the total
contribution of the last two equations.

\subsection{Hyperfine structure of states}
\label{22subsec:hfs}

\begin{figure}[t]
\begin{center}
\includegraphics[width=.47\textwidth]{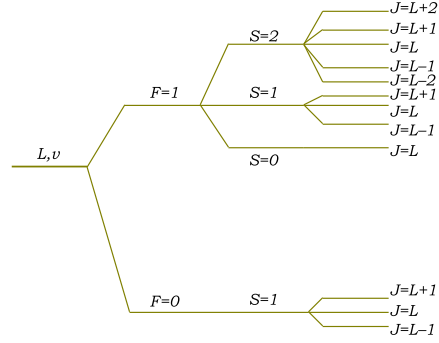}
\end{center}
\caption[]{Hyperfine structure of a ro-vibrational state of $\mbox{HD}^+$
with $L\ge2$.}
\label{22hfs-example}
\end{figure}

\index{Hydrogen molecular ions!hyperfine structure}
The leading order contribution to the hyperfine splitting of the
ro-vibrational states is calculated using the spin-dependent part of the
Breit-Pauli interaction Hamiltonian, with phenomenological values for the
nuclear magnetic moments and the electron anomalous magnetic moment. The
hyperfine levels of $\mbox{HD}^+$, $E_{vLFSJ}$, are labelled with the
quantum numbers $F$, $S$ and $J$ of the intermediate angular momenta
$\mathbf{F} = \mathbf{I}_p+\mathbf{s}_e$, $\mathbf{S}
=\mathbf{F}+\mathbf{I}_d$ and of the total angular momentum $\mathbf{J} =
\mathbf{L}+\mathbf{S}$ \cite{22HD_hfs}. In case of $\mbox{H}_2^+$ due to
Pauli exclusion principle the total nuclear spin $I$ is uniquely defined
by $L$ and parity of the electronic state. The following coupling scheme
is adopted: $\mathbf{F}=\mathbf{I}+\mathbf{s}_e$ and $\mathbf{J} =
\mathbf{L}+\mathbf{F}$ \cite{22H2_hfs}. The hyperfine structure (HFS) of the
ro-vibrational states of $\mbox{HD}^+$ consists of 4, 10 or 12 hyperfine
sub-levels for $L\!=\!0$, $L\!=\!1$ and $L\!\ge\!2$, respectively (see
Fig.~\ref{22hfs-example}). The multiplicity of the HFS of H$_2^+$ is reduced
to 1 for $L\!=\!0$, 5 for $L\!=\!1$, 2 for even, and 6 for odd $L$ states.
Typically, the hyperfine splitting of the lower ro-vibrational states of
HD$^+$ and H$_2^+$ is about 1 GHz. The uncertainty in the hyperfine
spectrum is related to the unknown contribution of the spin interaction
terms of orders $O(R_{\infty} \alpha^4(m/M))$ and higher, which have not
yet been taken into consideration, and is estimated not to exceed 100 kHz.

Each transition line between ro-vibrational states is split into a
multiplet of hyperfine components, corresponding to the allowed
transitions $i\!\to\!f$ between the states of the hyperfine structure of
the initial and final states. Whether these hyperfine components will be
resolved or the spectrum profile depends on the initial and final state
lifetime and on the experimental conditions (transition linewidth
$\Gamma_f$, laser intensity $I$, temperature, interaction time etc.)
Examples of spectral data are presented in experimental sections of our
review (see Fig.~\ref{22fig:results}). The shape of the profile also
depends on the population of the initial hyperfine states. The general
expressions for the probability per unit time for one- and two-photon
transitions between ro-vibrational states of the MHIs with account of the
hyperfine structure are given in Ref.~\cite{22exa}.

The probability per unit time for the hyperfine transition $i\!\to\!f$ at
resonance (averaged over the magnetic numbers of initial and final
states), $\Gamma_{f,i}$, may be represented in the form:
\begin{equation}
\label{22eq:intens}
\Gamma_{f,i}= T^2_{f,i}\,\Gamma_{v'L',vL},\qquad
\Gamma_{v'L',vL}=\frac{2\pi\alpha}{3\hbar}\,\frac{I}{\Gamma_f}\,
\frac{\langle v'L'||\,{\mathbf d}\,||vL\rangle^2}{2L+1}.
\end{equation}
Here $\Gamma_{v'L',vL}$ is the probability per unit time of
laser-stimulated dipole transitions between ro-vibrational states,
$\langle v'L'||\,{\mathbf d}\,||vL\rangle$ is the reduced matrix element
of the electric dipole moment of the ${\mathrm H}{\mathrm D}^+$ ion
$\mathbf{d}=\sum_a z_a \mathbf{r}_a$, and
\begin{equation}
T_{f,i}=\sqrt{(2J'+1)(2L+1)}
\sum\limits_{F''S''}(-1)^{S''+J+L'}
\left\{ \begin{array}{ccc}
{L} & {1} & {L'} \\ {J'} & {S''} & {J}
\end{array} \right\}
\beta^{f}_{F''S''}\beta^{i}_{F''S''},
\end{equation}
where $\beta^{vLFSJ}_{F''S''}$ are constant amplitudes of the state vectors
of the hyperfine states:
\begin{equation}
|vLFSJ,J_z\rangle=
   \sum\limits_{F''S''}\beta^{vLFSJ}_{F''S''}
   \sum\limits_{M\zeta} C_{LM,{S''}\zeta}^{JJ_z}
      \Psi_{vLM}({\mathbf R},{\mathbf r}_{1})\chi(F''S'',\zeta),
\label{22zeroth}
\end{equation}
determined from the effective Hamiltonian of spin interaction. Here
$\chi(FS,\zeta)$ are basis spinors of definite values of $F$, $S$ and
$S_z$ in the space of the spin variables. The relative intensity of the
hyperfine components of a transition line between ro-vibrational states is
thus determined by the amplitudes $T_{f,i}$. In case the individual
hyperfine components cannot be resolved, the observable intensity is
reduced to the intensity of the dipole ro-vibrational transition
$\Gamma_{v'L',vL}$, in agreement with the identity $\sum_f T^2_{f,i}=1$.
\index{Hydrogen molecular ions!intensities of transition lines}

\begin{figure}[t]
\begin{center}
\includegraphics*[width=.6\textwidth,height=37mm]{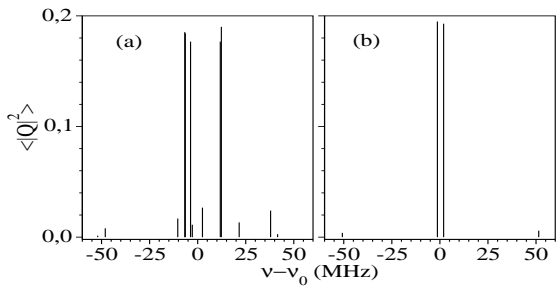}
\vspace{-3mm}
\end{center}
\caption[]{Hyperfine splitting and intensities of the two-photon
ro-vibrational transition line for the $\mbox{H}_2^+$ ion:
(a) $(v\!=\!0,L\!=\!1)\!\to\!(v'\!=\!1,L'\!=\!1)$,
(b) $(v\!=\!0,L\!=\!2)\!\to\!(v'\!=\!1,L'\!=\!2)$.}
\label{22HFS_laser}
\end{figure}

The hyperfine structure of the one- and two-photon transition lines
includes a large number of components, most of which, however, are
suppressed. There are as well dominant (or ``{\em favoured}'') transitions
between states with similar spin structure, such as
$(vLFJ)\rightarrow(v'L'FJ')$ with $\Delta J\!=\!\Delta L$ (for H$_2^+$).
In such pairs of homologous hyperfine states the spin-dependent
corrections to the ro-vibrational energies $E_{v'L'FJ'}$ and $E_{vLFJ}$
have close values, which partially cancel each other when evaluating the
spin correction to the resonance transition frequency
$(E_{v'L'FJ'}\!-\!E_{vLFJ})/h$. Indeed, the {\em favoured} hyperfine
transitions span over a frequency interval less than $25$ MHz (see
Fig~\ref{22HFS_laser}). It is natural to expect that the unknown
contributions to the frequency of the favoured transitions from the spin
interactions of order $R_{\infty}\alpha^4(m/M)$ and higher also tend to
cancel each other; therefore, the theoretical uncertainty of the resonance
frequency of the {\em favoured} hyperfine sublines will be less than
$\sim5$ kHz.

\subsection{Two-photon transition probabilities}
\label{22subsec:2photon}

\index{Doppler-free spectroscopy}
In order to assess the feasibility of Doppler-free two-photon spectroscopy
in H$_2^+$ or HD$^+$, it is essential to evaluate transition probabilities
between ro-vibrational states. This can be done using the formula from
second-order perturbation theory \cite{22Biraben73}, and the accurate
wavefunctions provided by variational calculations. Assuming that magnetic
sublevels are equally populated, the two-photon transition probability at
resonance between states $v,L$ and $v',L'$  is
\begin{equation}\label{22eq:rate}
\Gamma_{v,L,v',L'} =
   \left( \frac{4\pi a_{0}^{3}}{\hbar c} \right)^{2} \;
   \frac{4 I^{2}} {\Gamma_{f}} \overline{Q}_{v,L,v',L'}
\end{equation}
where $I$ is the excitation intensity, $\Gamma_{f}$ the transition
linewidth, and
\begin{equation}
\overline{Q}_{v,L,v',L'} =
  \frac{1}{2L+1} \; \sum_{k=0,2} \;
     \frac{\left| \langle v L \| Q^{(k)} \| v' L' \rangle \right|^2} {2k+1}.
\end{equation}
$Q^{(0)}$, $Q^{(2)}$ are respectively the scalar and tensor parts of the
two-photon transition operator
\begin{equation}
Q = \frac{1}{4 \pi \epsilon_{0} a_{0}^{3}} \;
   \mathbf{d}\!\cdot\!\mbox{\boldmath$\epsilon$} \; \frac{1}{E-H} \;
   \mathbf{d}\!\cdot\!\mbox{\boldmath$\epsilon$}
\end{equation}
Here, $E = [E(v,L)+E(v',L')]/2$ is the one-photon resonance energy and
{\boldmath$\epsilon$} the exciting field polarization.
The two-photon transition probabilities were calculated in
\cite{22paris2,22paris_JPB01}.

For the H$_2^+$ case \cite{22paris_JPB01} there exists a quasi-selection
rule $\Delta v = \pm 1$, and the dimensionless transition probabilities
$\overline{Q}_{v,L,v',L'}$ are rather small, of the order of 1, which is
due to the level structure of H$_2^+$. If we consider the example of
transitions between $L=0$ states, these states are of $^{1}S^e$ symmetry,
and there is no resonant intermediate level of $^{1}P^o$ symmetry that
could enhance the transition probability.

The situation is different in the HD$^+$ case \cite{22paris2}, since there
is no splitting between singlet and triplet state due to the loss of
exchange symmetry between the nuclei. As a result, for a transition
between $L\!=\!0$ (S$^e$) states, there exist intermediate bound P$^o$
levels which can be very close in energy and efficiently enhance the
transition probability. This is most likely to happen if the difference
between $v$ and $v'$ (the initial and final vibrational quantum numbers)
is an even number. In this case, the state
$(v''\!=\!(v\!+\!v')/2,L''\!=\!L\!\pm\!1)$ is often close to the middle
energy $E = [E(v,L)+E(v',L')]/2$, because of the quasi-harmonic structure
of vibrational levels. As a result, some of the most intense two-photon
lines are $\Delta v\!=\!2$ transitions in the 5-6 $\mu$m range or $\Delta
v\!=\!4$ transitions in the 2.5-3 $\mu$m range, accessible e.g.\ with
continuous-wave optical parametric oscillator or quantum cascade lasers.
The dimensionless transition probabilities $\overline{Q}_{v,L,v',L'}$ can
reach values as high as 300 for the $(v\!=\!0,L\!=\!1) \to
(v\!=\!2,L\!=\!1)$ transition at 5.366 $\mu$m. Thus, the HD$^+$ molecular
ion is a promising candidate for precise two-photon spectroscopy.

The next step is to consider the hyperfine structure of two-photon
transition lines. The representation of the hyperfine state vectors in
the H$_2^+$ case is
\begin{equation}
|vLFJJ_z \rangle = \sum_{F'} \; \beta_{F'}^{vLFJ} \; \sum_{M \zeta} \; C_{LM,F'\zeta}^{JJ_z} \; \psi_{vLM}
(\mathbf{R,r_1}) \; \chi(F',\zeta).
\end{equation}
The definitions for $\chi(F',\zeta)$ and $\beta_{F'}^{vLFJ}$ are similar
to those in (\ref{22zeroth}). Assuming equal populations for
hyperfine magnetic sublevels, the dimensionless transition probability
between levels $|i \rangle = |vLFJ \rangle$ and $|f \rangle =
|v'L'F'J'\rangle$ is:
\begin{equation}
\hspace{0.5mm}
\overline{Q}_{i,f}\! =\!
   (2J'\!+\!1)\! \sum_{k=0,2}\!
      \frac{\left| \langle v L \| Q^{(k)} \| v' L' \rangle %\right|^2
      %\left|
      \sum_{\scriptscriptstyle F''}^{} (-1)^{J'\!+\!L\!+\!F''}\!
         \left\{\!
            \begin{array}{ccc}
               L\! & k\! & L' \\
               J'\! & F''\! & J
            \end{array}\!
         \right\} \beta_{F''}^f \beta_{F''}^i\right|^2\!\!} {2k+1}
\hspace{-3mm}
\end{equation}
\index{Hydrogen molecular ions!intensities of transition lines}
\index{H2plus@H$_2^+$ ions!intensities of HFS lines}
The hyperfine structure of two different transitions for linear
polarization is shown in Fig.~\ref{22HFS_laser}. Transitions between odd
$L$ states (Fig.~\ref{22HFS_laser}(a)) comprise between 25 to 34
components, 5 or 6 of which are favoured. Transitions between even $L$
states (Fig.~\ref{22HFS_laser}(b)) have a much simpler structure, the
total nuclear spin being zero. The non zero even $L$ spectrum comprises
only two main components (verifying $\Delta J\!=\!0$) together with two
weak satellites. Transitions between $L\!=\!0$ states are structureless,
which makes them especially attractive from a metrological point of view.

%------------------------------------------------------------------
\section{Two-photon spectroscopy of H$_2^+$}

\index{H2plus@H$_2^+$ ions}

A two-photon vibrational spectroscopy experiment aimed at the
determination of the electron to proton mass ratio is being setup at the
Kastler Brossel Laboratory. We begin by recalling the basic spectral
features of the MHI, and discuss the planned experimental sequence. In the
second part, we report on the present status of the experimental setup. It
is composed of a hyperbolic Paul trap in which a few thousand H$_2^+$ ions
can be confined, a UV laser for ion preparation and detection by
state-selective photodissociation, and a narrow-line, tunable laser system
that will excite the two-photon transition.

\subsection{H$_2^+$ level structure}

\begin{figure}[t]
\begin{center}
\hspace{-5mm}\includegraphics*[width=75mm,height=48mm]{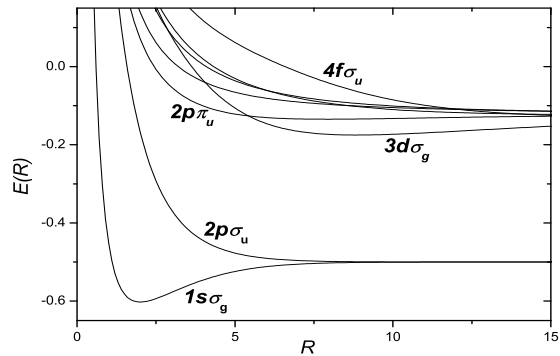}
\end{center}
\vspace{-4mm}
\caption{\label{22B-O}Born-Oppenheimer electronic energies (in a.u.) of the
adiabatic potential terms.}
\end{figure}

\index{Born-Oppenheimer approximation}
\index{Hydrogen molecular ions!level structure}
Although the Born Oppenheimer (BO) approximation is not relevant for
highly accurate calculations, it remains a very convenient tool to get a
useful insight into the H$_2^+$ level structure. In order to understand
the processes discussed here, it is enough to consider the first two BO
electronic curves: the ground state $1s\sigma_g$ and first excited state
$2p\sigma_u$, which are depicted in Fig.~\ref{22B-O}.

The exact symmetries of the system are the total spatial parity $\pi$ and
the exchange of nuclei $P_{12}$; the $g/u$ electronic parity $\pi_e$ used
in the BO approach is related to them by $\pi_e=\pi P_{12}$. The bound
levels of H$_2^+$ can be labelled $v,(^{2I+1}L^{e,o})$ where $v$ and $L$
are the vibrational and orbital quantum numbers, $I$ is the total nuclear
spin quantum number and $(e,o)$ stands for the total parity.
Since the total spatial parity is $\pi=(-1)^L$, the $1s\sigma_g$ curve
only supports $^1S^e$, $^3P^o$, $^1D^e\dots$ levels.

The $2p\sigma_u$ electronic curve presents at large internuclear distances
a weak attractive potential that supports two bound $L=0$ energy levels
\cite{22moss-H2+,22weak}. Some of those states have been observed by microwave
or laser spectroscopy experiments \cite{22Carr93}. At higher
$L$ the $2p\sigma_u$ potential supports bound or dissociative $^1P^o$,
$^3D^e$ $^1F^o\dots$~states that can be calculated numerically using
either the variational, or the complex coordinate rotation method.

\subsection{One-photon transitions: photodissociation}

\begin{figure}[t]
\begin{center}
\includegraphics*[width=47mm,height=75mm,angle=-90]{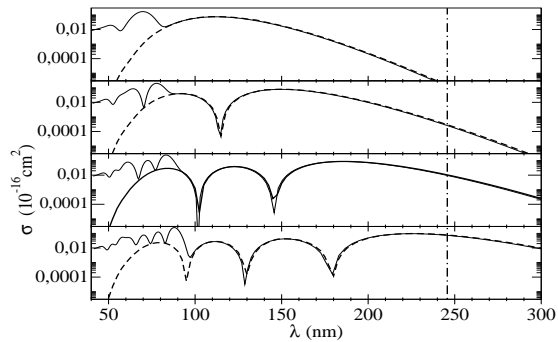}
\end{center}
\vspace{-5mm}
\caption{\label{22photodiss}
Photodissociation cross sections of the $L\!=\!0$, $v$ levels of H$_2^+$. The
dashed lines are the result of a Born Oppenheimer calculation~\cite{22dunn},
which takes the $1s\sigma_g$ and $2p\sigma_u$ electronic curves into account.
The nodal structure of the cross section reflects that of the
vibrational wavefunction. The solid lines are obtained from an exact
variational method which fully takes into account the three body
dynamics~\cite{22kilic}. The additional nodal structure appearing in the
short wavelength domain can be interpreted as the
photodissociation to higher excited electronic states ($3d\sigma_g$,
$2p\pi_u$, $4f\sigma_u$, \dots). The dot-dashed vertical line
corresponds to the KrF laser wavelength of 248~nm.}
\end{figure}

\index{H2plus@H$_2^+$ ions!photodissociation}
The selection rules for one-photon dipole transitions are $\Delta L= \pm
1$ and $\Delta I=0$. As a consequence, transitions between bound
ro-vibrational states of H$_2^+$ are forbidden (in contrast with the
HD$^+$ case), resulting in very long-lived states. On the other hand,
one-photon photodissociation transitions from $1s\sigma_g$ to $2p\sigma_u$
electronic states are allowed. The photodissociation cross sections
$\sigma_v$ of the $(^1S^e,v)$ states have been first computed by
Dunn~\cite{22dunn} in the Born-Oppenheimer approximation, and then using the
perimetric coordinate variational method in~\cite{22kilic}. The results are
given in Fig.~\ref{22photodiss}. They show that a laser source in the
250~nm range can be used to selectively photodissociate the
$v=1,2,3,\dots~$ vibrational states while keeping the ions in the $v=0$
level since the successive cross section ratios $\sigma_{v+1}/\sigma_v$
are 214, 40, 10, for $v=0,1,2$, respectively.

\subsection{Two-photon transitions: choice of the transition}

\index{Doppler-free spectroscopy}
One-photon transitions between bound states being forbidden, a high
resolution study of the vibrational structure of H$_2^+$ is only possible
using Doppler-free two-photon spectroscopy. Two-photon transitions obey
the selection rule $\Delta L=0,\pm 2$ as well as the quasi selection rule
$\Delta v = \pm 1$, as discussed in section \ref{22subsec:2photon}. The
corresponding $(v,L) \to (v'\!=\!v\!+\!1,L')$ transition frequencies lie
in the 8-12~$\mu$m range. Among them, we have chosen to probe the
$v\!=\!0\to v\!=\!1$ transitions, for $L$ and $L'$ equal to 0 or 2 and
eventually 1 or 3. We now give the arguments that explain this choice.

The first condition to fulfill is that it should be possible to prepare a
large enough number of H$_2^+$ in the initial state of the transition. The
ro-vibrational populations of H$_2^+$ ions, after creation by electron
impact ionization of a low pressure H$_2$ gas, have been studied both
theoretically and experimentally~\cite{22werth-H2+}. The vibrational
populations are linked to the overlap of the H$_2$ and H$_2^+$ vibrational
wave functions (Franck-Condon principle); they are found to be of the
order of 12, 18, 18, 15, 11, 8, 5, 4\% for the first few levels. The
rotational populations of the H$_2^+$ ions are those of the H$_2$ mother
molecules, i.e. 12, 28, 28, 18, 8\% at 300 K. Moreover, we have shown in
the previous paragraph that UV photodissociation provides a convenient way
to prepare ions in the ground vibrational state; it is then desirable to
choose a $v\!=\!0$ state as initial state of the transition, with $L$
between 0 and 3, $L\!=\!1$ or 2 being the best choices with respect to the
number of ions. The same photodissociation process can be used to detect
the ions in the excited $v\!=\!1$ state.

The hyperfine structure of two-photon lines should also be considered; it
is apparent from Fig.~\ref{22HFS_laser} that it is simpler for transitions
between even $L$ states. Interpretation of experimental data is likely to
be easier in this case.

The intensities of the various $(v\!=\!0,L) \to (v'\!=\!1,L')$ (with low
$L$) two-photon lines are of the same magnitude; the choice of a
particular transition depends mostly on the availability and
characteristics of laser sources at the required wavelength. The whole
mid-infrared range is accessible by the recently developed quantum cascade
lasers (QCL); $L'\!=\!L$ transitions are especially attractive, because
they lie within the spectrum of CO$_{2}$ lasers ($\lambda \simeq$ 9-10
$\mu$m). Also, a number of frequency reference molecular absorption lines
are known in this range~\cite{22hitran}. The first transition that is going
to be probed in our experiment is the $(v\!=\!0,L\!=\!2) \to
(v'\!=\!1,L'\!=\!2)$ line at 9.166 nm. The details of coincidences with
CO$_2$ lines and molecular reference lines, which make this transition
favorable, are explained below.

\subsection{Experimental sequence}

The two-photon transition matrix elements $|Q_{v,L,v',L'}|^2$ of the
"favoured" hyperfine components of two-photon transitions are of the order
of 0.2 (see Fig.~\ref{22HFS_laser}). A typical QCL can deliver about 50~mW
of single-mode optical power. Assuming a perfect coupling to a build-up
cavity of finesse 1000 with a waist of 1~mm, one obtains a laser flux of
15~W/mm$^2$. Assuming an instrumental width $\Gamma_f \approx$10~kHz,
equation (\ref{22eq:rate}) yields transition rates of about 70/s. This order
of magnitude shows that long interaction times are needed and that one has
to work with a cloud of trapped ions, having a radius of the order of the
beam waist.

\index{Resonance Enhanced Multiphoton Dissociation (REMPD) method}
The considerations of the previous paragraph show that vibrational
two-photon spectroscopy of H$_2^+$ can be performed by ($2\!+\!1'$)
resonance enhanced multiphoton dissociation (REMPD). This process is very
similar to that implemented for HD$^+$ vibrational spectroscopy and
described in more detail in Section~\ref{22sec:HD+}.

The experiment will be conducted in the following stages:
\begin{itemize}
\item simultaneous creation, trapping and selection of $(L,v=0)$ H$_2^+$ ions.
\item excitation of the $(L,v\!=\!0)\to(L',v'\!=\!1)$ two-photon transition.
\item photodissociation of the $(L,v\!=\!1)$ H$_2^+$ ions.
\item time of flight detection of H$^+$ and H$_2^+$ ions.
\end{itemize}

\subsection{Experimental setup}

\begin{figure}[t]
\begin{center}
\hspace{-4mm}
\hbox{(a)
%\raisebox{20mm}{\includegraphics*[width=30mm,angle=-90]{22p_ion_trap.eps}}}
\raisebox{-11mm}{\includegraphics*[width=50mm]{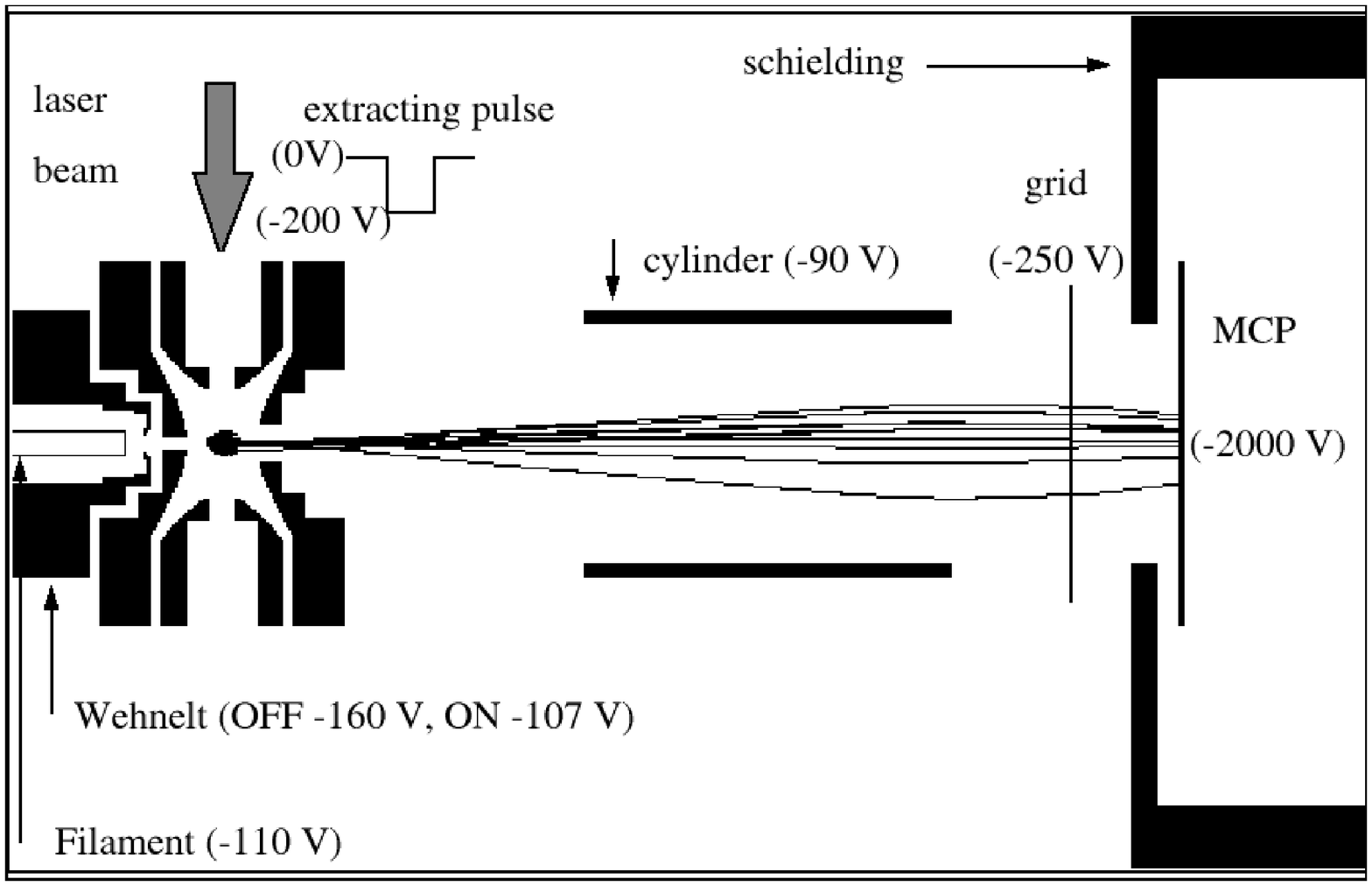}}}
\hspace{8mm}
\hbox{(b)\hspace{-3mm}
\raisebox{-14mm}{\includegraphics*[width=5cm]{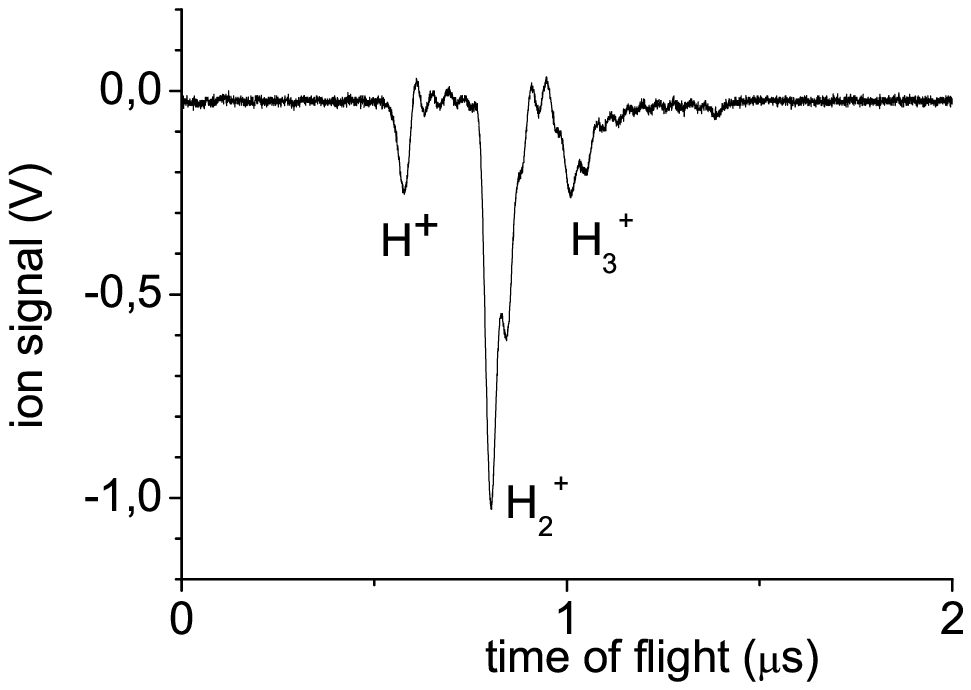}}}
\end{center}
\vspace{-5mm}
\caption{\label{22ion_trap}
(a) Simulation of the experimental setup for ion creation, trapping and
    detection using the SIMION7 software. MCP is a multichannel plate
    detector.
(b) Time of flight spectrum showing the H$^+$, H$_2^+$ and H$_3^+$
    species confined in the Paul trap.
}
\end{figure}

%\subsubsection{Ion trap}

\index{H2plus@H$_2^+$ ions!ion trap}
The ion trap is depicted in Fig.~\ref{22ion_trap}(a). It is a hyperbolic
Paul trap with a ring of inner radius $r_0 = 8.2$~mm and two end caps
separated by $2z_0 = 6$~mm. Two pairs of holes (2~mm in diameter) are
drilled in the ring along two orthogonal axes to shine the ion cloud with
the UV and IR light. Both end cap electrodes  are AC grounded. A RF
voltage (about 200~V peak to peak amplitude at 10.3~MHz) and a continuous
voltage of a few Volts are applied to the ring electrode, resulting in
trapping well depths of a few eV.

The H$_2^+$ ions are produced by electron impact ionisation from the
residual H$_2$ gas. The electron gun is made of a tungsten wire and a
Wehnelt cylinder; it is typically turned on for 100-200~ms. A 1~mm hole in
one of the end cap electrodes allows access to the trap.

The contents of the trap are analyzed by applying a short negative high
voltage pulse to the second end cap, thus extracting the ions from the
trap through a 2~mm hole. The extracted ions are accelerated and focused
onto a multi-channel plate (MCP) detector located 7~cm away, a long enough
distance to separate by time of flight the H$^+$, H$_2^+$ and H$_3^+$ ions
that are simultaneously produced and trapped. A typical time of flight
spectrum is shown in Fig.~\ref{22ion_trap}(b). Up to a few thousand
H$_2^+$ ions can be stored in the trap. The ion lifetime is of a few
seconds and is limited by the residual pressure in the vacuum chamber.

The undesirable H$^+$ and H$_3^+$ ions are eliminated using the parametric
excitation of their secular motion, by applying RF voltage in the MHz
range on one of the end cap electrodes during the ionisation process. A
KrF excimer laser at 248~nm is used to photodissociate the $v\!\geq\!1$
states in order to produce a $(L,v\!=\!0)$ ions cloud. The ions are shined
by 20~mJ pulses during the filling of the trap. The characterization of
ro-vibrational populations of the resulting ion cloud is now in progress.

Since all the bound states of H$_2^+$ are metastable, the natural widths
of the two-photon transitions are extremely small. In Paul traps, the ion
cloud temperature is of the order of magnitude of the potential depth
expressed in K, i.e.\ $\approx\!10^4$~K in our trap. Under those
conditions, the two-photon linewidth $\Gamma_f$ (appearing in
Eq.~(\ref{22eq:rate})) is expected to be limited by the second-order Doppler
effect, i.e.\ of the order 10~kHz. It will limit the ultimate frequency
resolution of the experiment at the $3\!\cdot\!10^{-10}$ level, and the
mass ratio resolution at the $6\!\cdot\!10^{-10}$ level.

\begin{figure}[t]
%\vspace{-2mm}
\begin{center}
\includegraphics*[width=47mm,angle=-90]{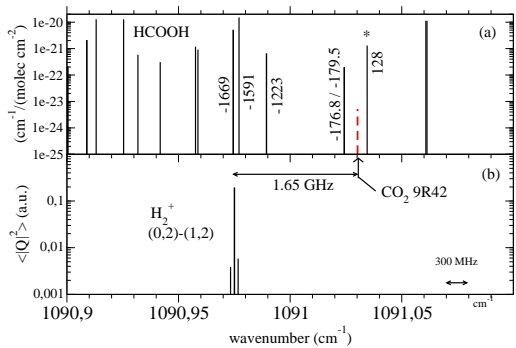}
\end{center}
\vspace{-3mm}
\caption{\label{22HCOOHspectra}
(a) Absorption spectrum of formic acid (HCOOH)~\cite{22hitran}. The line
    intensities are given in cm$^{-1}$/(molecule cm$^{-2}$).
(b) Two-photon transition probabilities in atomic units. The central peak is
    made of two close components (see Fig.~\ref{22HFS_laser}). The
    dashed line is the 9R(42) CO$_2$ emission line. The detunings between the
    9R(42) CO$_2$ line and the HCOOH lines are indicated in MHz.
    The CO$_2$ laser is locked to the HCOOH line indicated by the star.}
\end{figure}

Ion cooling will thus be necessary in order to reach the metrological
objective of the experiment at the 10$^{-10}$ level. Nevertheless, the
first step of the experiment is the observation of a two-photon
transition, which is feasible with hot ions using a kHz linewidth laser
source.

\subsection{Two-photon excitation laser source}

\begin{figure}[t]
\begin{center}
\includegraphics[width=8cm]{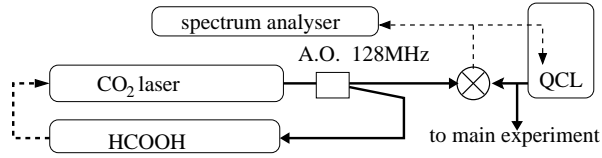}
\end{center}
\caption{\label{22QCL}Simplified setup of the CO$_2$/HCOOH phase-locked
quantum cascade laser source. The QCL is mounted in a liquid nitrogen
optical cryostat. The mixer is a room temperature HgCdZnTe detector.
Solid lines are optical paths. Dashed lines are electrical paths and servo
loops. A.O.\ is an acousto-optic modulator.}
\end{figure}

\index{H2plus@H$_2^+$ ions!laser system}
The laser system we have built is aimed at exciting the $(L\!=\!2,
v\!=\!0) \to (L'\!=\!2,v'\!=\!1)$ two-photon transition at 9.166 $\mu$m.
In this range, two kinds of laser sources are available. Single-mode
CO$_2$ lasers have high output power and sub-kHz linewidths, but are
hardly tunable on ranges exceeding 100~MHz, i.e. much smaller than the
1.65~GHz gap between the closest CO$_2$ emission line (9R(42)) and the
H$_2^+$ line (see Fig.~\ref{22HCOOHspectra}(b)). Recently, single mode
quantum cascade laser (QCL) became commercially available. They can be
tuned over about 10~cm$^{-1}$ (300~GHz) through their temperature and
injection current, but have large linewidths of the order of a few MHz.
Several experiments have shown that the linewidth can be reduced well
below the kHz level by injection-current locking the QCL to a molecular
line~\cite{22QCL-molec} or to a high finesse Fabry-Perot cavity
resonance~\cite{22QCL-cav}. We have developed a laser source that takes
advantage of both the narrow linewidth of the CO$_2$ laser and the
tunability of the QCL \cite{22bielsa-QCL}.

\begin{figure}[b]
\begin{center}
\hspace{-6mm}\includegraphics*[width=45mm,angle=-90]{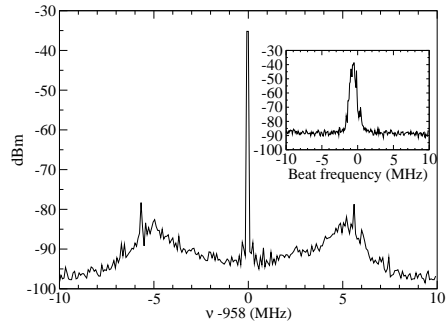}
\end{center}
\vspace{-4mm}
\caption{\label{22phaselock}
Phase-locked beat note between the QCL and the CO$_2$/HCOOH frequency
reference. RBW 10 kHz, VBW 1 kHz. The spectrum shows a loop bandwidth of
the order of  6 MHz. The central peak is extremely narrow, with a -3dB
width smaller than the 200 Hz resolution of the spectrum analyser. The
inset shows the free running beat note with the same scale and a 500 kHz
RBW.}
\end{figure}

The setup is shown in Fig.~\ref{22QCL}. A CO$_2$ laser oscillating on the
9R(42) line is frequency shifted by 128~MHz and stabilized on the
intracavity saturated absorption signal of the $(21,3,19)\!\to\!(21,2,20)$
line of the $\nu_6$ band of formic acid (HCOOH) (see
Fig.~\ref{22HCOOHspectra}(a)). The absolute frequency of that transition
(32 708 263 980.5 kHz) has recently be determined with an uncertainty of
1~kHz \cite{22c56} by sum frequency mixing with a 30 THz wide
visible femtosecond frequency comb \cite{22c57}. The QCL is operated in a liquid
nitrogen optical cryostat. The output power is 50~mW with a 700~mA
injection current and a temperature of 80K. The QCL is phase-locked to the
CO$_2$ laser with a tunable frequency offset in the 500-2000~MHz range
\cite{22bielsa-QCL}. The analysis of the beat note spectrum under locked
conditions (see Fig.~\ref{22phaselock}) shows that we have realized a
narrow-line tunable laser source well suited to probe the H$_2^+$
two-photon lines, and also the ro-vibrationnal spectrum of
HCOOH~\cite{22c57} or other molecules (NH$_3$, \dots) of atmospheric or
astrophysical interest.

%------------------------------------------------------------------------
\section{Cooling and Spectroscopy of HD$^+$}
\label{22sec:HD+}

In experiments performed at the University of D\"usseldorf, the MHIs
H$_2^+$, D$_2^+$, and HD$^+$ were cooled to temperatures of $\simeq10\,$
mK in a radiofrequency trap, by sympathetic cooling with laser-cooled
beryllium ions. High-resolution spectroscopic studies of several
rovibrational infrared transitions in HD$^+$ were performed. Hyperfine
splitting of the lines was observed, and is in good agreement with
theoretical predictions. The transitions were detected by monitoring the
decrease in ion number after selective photodissociation of HD$^+$ ions in
the upper vibrational state.

\subsection{Preparation and characterization of cold MHI ensembles}

\index{HDplus@HD$^+$ ions!sympathetic cooling}
MHIs are just a few of a multitude of species that can be cooled
to mK temperatures, by sympathetic cooling
\cite{22Larson1986,22Drewsen2000} - the molecular species and a
laser-coolable atomic species, with the same sign of charge, are
simultaneously stored in a radiofrequency trap. Laser cooling the
atoms then also efficiently cools the molecular ions via the
long-range Coulomb interaction. Temperatures below 20 mK can be
reliably reached. We have shown that using Be$^+$ ions as coolant
permits to cool sympathetically ions from mass 1 to mass 200 amu
\cite{22Blythe2005,22RothJPhysB2006,22RothPRA2006}. A heavier atomic
coolant species can be used to extend the mass range. For example,
using $^{138}$Ba$^+$ as coolant molecular ions up to mass 410 amu
have recently been cooled \cite{22Ostendorf2006}.

We use a linear quadrupole trap to simultaneously store both Be$^+$ and
MHIs. The radiofrequency trap is driven at 14~MHz, with a peak-to-peak
amplitude of 380~V.  This results in a radial Mathieu stability parameter
$q_r \simeq 0.13$ for HD$^+$. The trap is enclosed in a UHV chamber kept
below $10^{-10}$~mbar. The chamber is equipped with a leak valve for the
controlled introduction of gases. An all-solid-state 313~nm laser system
is used for cooling Be$^+$ \cite{22Schnitzler2002}.

\begin{figure}[tbp]
  \begin{center}
    \includegraphics[width=0.6\textwidth]{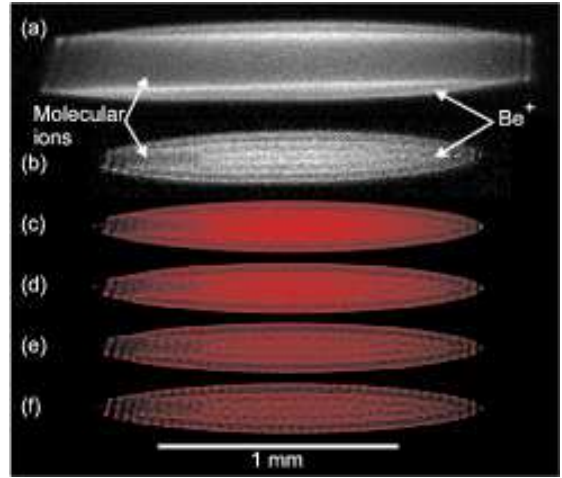}
  \end{center}
    \caption{\label{22fig:hdproduction} Fluorescence images of
    (a) a large ion crystal with a high fraction of sympathetically cooled
    ions (approximately 1200 light ions and 800 Be$^+$ ions)
    (b) a smaller crystal containing approx.\ 690 Be$^+$ ions, and 12 (exactly)
    HD$^+$ ions, and simulated images of this crystal at
    (c) 20~mK, (d) 12~mK, (e) 8~mK and (f) 6~mK.
    In the simulations, the beryllium ions are shown in red and the HD$^+$
    ions in blue. Laser cooling beam propagation is to the right, along the
    $z$-axis \cite{22Blythe2005}.}
\end{figure}

To load Be$^+$ ions into the trap, atoms are thermally evaporated from a
beryllium wire, and ionised by an electron beam. The molecular
loading is achieved by leaking in neutral gas at a pressure of $\sim\!
(1\!-\!3)\cdot\! 10^{-10}$~mbar, ionised by an electron beam with an
energy of 200~eV, and a current of $\sim 30~\mathrm{\mu A}$, for a loading
time of 2~s. This produces mixed-species crystals like those shown in
Figs.~\ref{22fig:hdproduction}(a,b). The ions with a higher charge-to-mass
ratio (in this case the molecular ions) experience a stronger trap
pseudopotential, and thus form a dark (non-fluorescing) core to the
crystal. The asymmetric distribution of species along the $z$-axis
observed in Fig.~\ref{22fig:hdproduction}(b) is caused by the light pressure
of the cooling laser on the beryllium ions.

The observed crystals are reproduced by molecular dynamics (MD)
simulations \cite{22Blythe2005,22RothBa2005}. Visual matching of overall
structure, structural details and blurrings of CCD and simulated images
allows fitting the ion numbers and temperatures of the different species.
The number of ions of different species given in
Fig.~\ref{22fig:hdproduction}(b) were found in this way. In the simulations
we assume an ideal linear trap, use the quasipotential approximation and
model heating effects by stochastic forces on the ions. The obtained
temperatures are thus effective secular temperatures.
Fig.~\ref{22fig:hdproduction} shows a determination of the temperature;
agreement is found for a Be$^+$ temperature of appox.\ 10~mK. This sets an
upper limit, as our experimental images are also limited in sharpness by
our detection optics, CCD resolution, and sensor noise, which are not
considered. The temperature varies depending on crystal size and cooling
parameters, and is typically in the range 5~mK to 15~mK, with smaller
crystals generally colder. These temperatures are consistent with
measurements of the fluorescence lineshape of the Be$^+$ ions.

For all species of molecular ions studied here, our MD simulations show
that the sympathetically cooled molecular ion ensemble is also
crystalline, i.e. its time-averaged ion distribution is strongly
inhomogeneous, and that it is strongly thermally coupled to the Be$^+$
ions. Assuming similar heating effects for the molecular ions and the
Be$^+$ ions, the simulations show that the moleculars ions have a
temperature similar to that of Be$^+$, due to the strong Coulomb coupling.

The trapped species are identified and the time evolution of their numbers
is monitored by excitation of their mass-dependent radial (secular) modes,
using a spatially homogenous and temporally oscillating electric field.
For HD$^+$ ions the measured secular frequency was $\approx$770\,kHz,
significantly shifted from the calculated single-particle frequency, due
to Coulomb coupling between different species in the trap
\cite{22Lineshifts2006}. Excitation amplitude, sweep rate, and covered
frequency range were chosen so that the ion crystal had sufficient time to
cool back to its initial temperature between individual excitation cycles.
The excitation heats both the molecular ions and the  atomic coolants,
which changes the scattering rate of 313 nm cooling light by the Be$^+$
ions. The HD$^+$ secular resonance becomes visible in the Be$^+$
fluorescence, and its strength is proportional to the amount of HD$^+$
ions in the ion crystal.

\subsection{Spectroscopy of HD$^+$}
\index{HDplus@HD$^+$ ions}
\begin{figure}[t]
  \begin{center}
  \includegraphics[width=7cm]{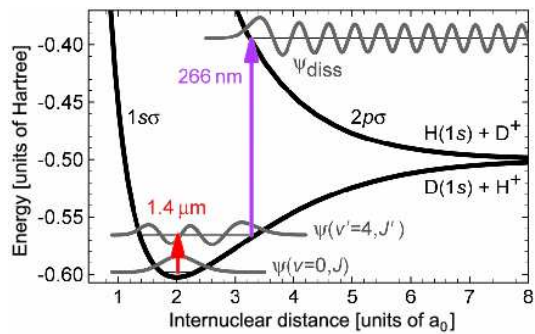}
  \end{center}
    \caption{\label{22fig:scheme} Principle of $(1\!+\!1')$ REMPD spectroscopy
    of HD$^+$ ions. A tunable IR diode laser excites a rovibrational overtone
    transition $(v\!=\!0,L)\!\to\!(v'\!=\!4,L')$.
    The HD$^+$ ions excited to the $v'\!=\!4$ vibrational level are
    dissociated using cw 266 nm laser radiation:
    $\mbox{HD}^+(v'\!=\!4)+h\nu\to\mbox{H}+\mbox{D}^+$, or
    $\mbox{H}^+ +\mbox{D}$.
    Due to different Franck-Condon wavefunction overlap, the calculated UV
    absorption cross section from the $v'\!=\!4$ level
    ($\sim\!2.4\!\times\!10^{-17}$ cm$^2$) is about 7 orders of magnitude
    larger than from $v\!=\!0$ \cite{22Tadjedine1977}.
    Energy values represent total binding energies of the molecule \cite{22Schiller06}.}
\end{figure}

\index{Resonance Enhanced Multiphoton Dissociation (REMPD) method}
The choice of HD$^+$ for spectroscopic studies was made because of the
availability of dipole-allowed ro-vibrational transitions which simplify
the spectroscopic techniques. Nevertheless, vibrational spectroscopy in
the electronic ground state in near-absence of collisions, as is the case
for the present molecular ions ensembles, is faced with the difficulty
that molecules excited to a vibrational level decay only slowly, implying
very low fluorescence rates. As the fluorescence wavelengths are in the
mid to far infrared, photon counting would require a sophisticated
detection system. We circumvent this difficulty by applying the technique
of ($1\!+\!1'$) resonance enhanced multiphoton dissociation (REMPD): the
molecules are excited by an infrared (IR) laser and then selectively
photodissociated from the upper vibrational state by a second,
fixed-wavelength ultraviolet (UV) laser (Fig.\ref{22fig:scheme}). The
remaining  number of molecular ions is the quantity measured as a function
of the frequency of the IR laser. As the molecular sample is small (typ.
40-100 ions) the spectroscopy requires the spectra to be obtained by
repeated molecular ion production and interrogation cycles. The lasers
employed are a single-frequency, widely tunable diode laser at 1.4 $\mu$m
(Agilent 81480A), and a resonantly frequency-doubled Yb:YAG laser at 266
nm. The IR laser linewidth was $\sim\!5$ MHz, and its frequency was
calibrated with an accuracy of 40 MHz by absorption spectroscopy in a
water vapor cell.

\index{HDplus@HD$^+$ ions!population of ro-vibrational states}
Due to the weak coupling between external and internal (rotational)
degrees of freedom, the internal temperature of the HD$^+$ ions is close
to room temperature, in thermal equilibrium with the vacuum chamber
\cite{22Bertelsen2006,22Koelemeij06}. There is significant ($>\!5\%$)
population for rotational levels up to $L\!=\!6$. Indeed, we have observed 12
transitions between 1391 nm and 1471 nm, from lower rotational levels
$L\!=\!0$ to $L\!=\!6$.

\begin{figure}[t]
  \begin{center}
  \includegraphics[width=0.45\textwidth]{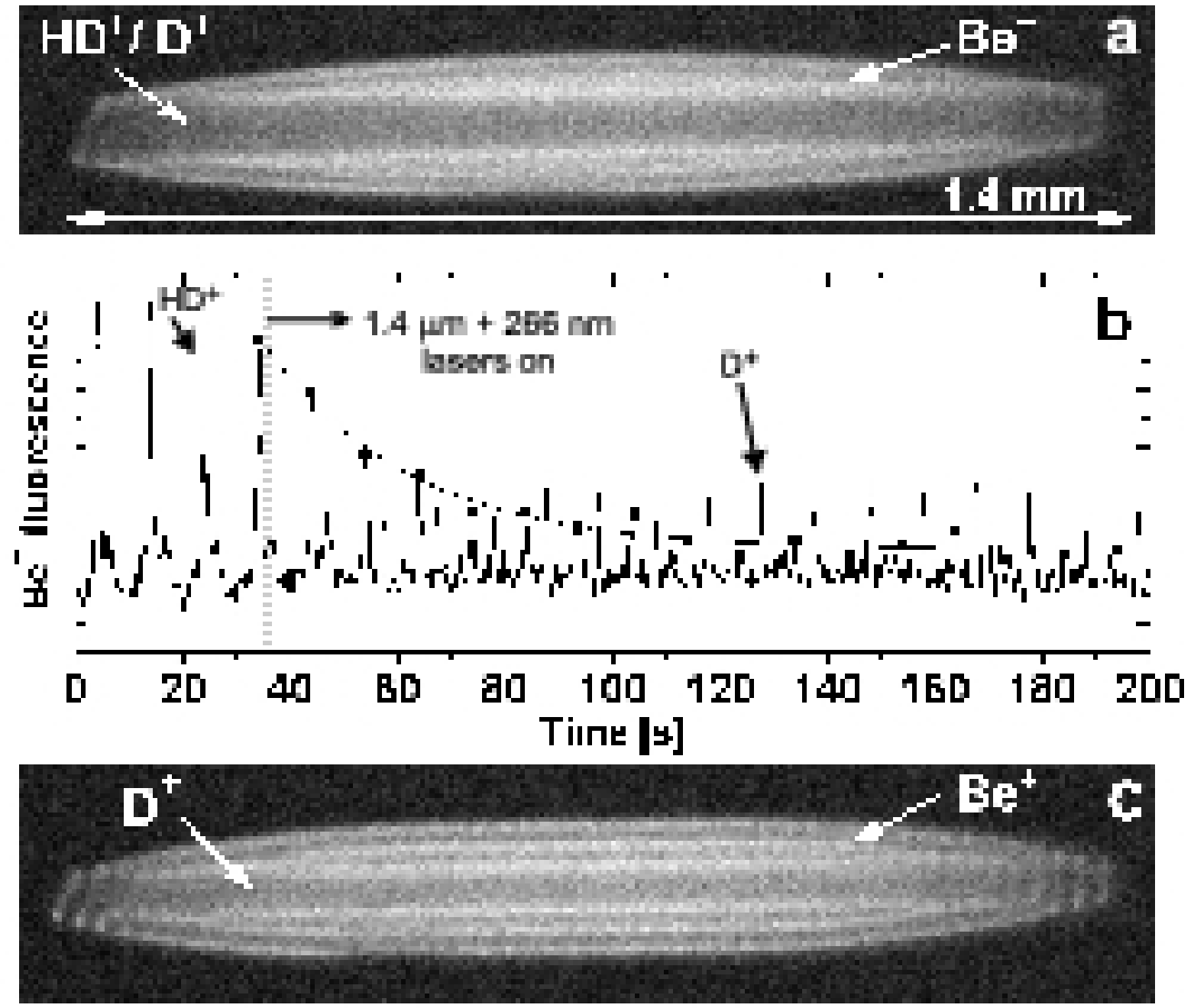}
  \raisebox{5mm}{
  \includegraphics[width=0.45\textwidth]{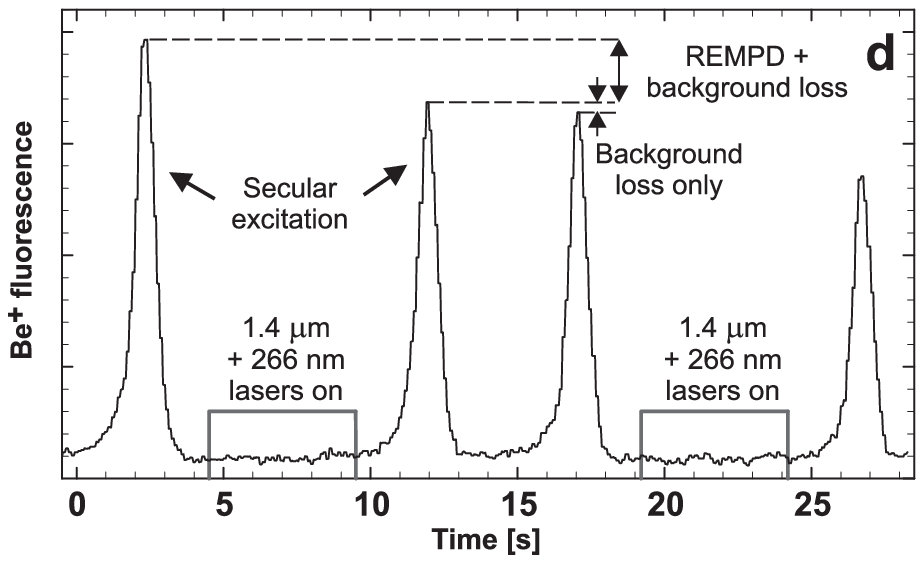}
  }
  \end{center}
    \caption{\label{22fig:method}
    (a) Initial ion crystal: $\approx$1100 Be$^+$, $\approx$100 HD$^+$,
    and $\approx$20 D$^+$ ions at $\approx$20 mK (the presence of cold
    HD$^+$ ions is obvious from the dark crystal core).
    (b) Repeated secular excitation of the crystal in (a) at 3 V amplitude.
    The excitation frequency was swept between 500 kHz and 1500 kHz.
    The IR laser is tuned to the maximum of the
    $(v\!=\!0,L\!=\!2)\to(v'\!=\!4,L'\!=\!1)$ line.
    The curve is an exponential fit with a decay constant of 0.04 s$^{-1}$.
    (c) Ion crystal after dissociation of all HD$^+$ ions:
    $\approx$1100 Be$^+$ and $\approx$50 D$^+$ ions at $\approx$20 mK.
    (d) Measurement cycle consisting of repeated probing of the number of
    HD$^+$ ions before and after exposure to the spectroscopy lasers
    \cite{22Schiller06}.
    }
\end{figure}

\begin{figure}[t]
\begin{center}
    \includegraphics[width=0.98\textwidth]{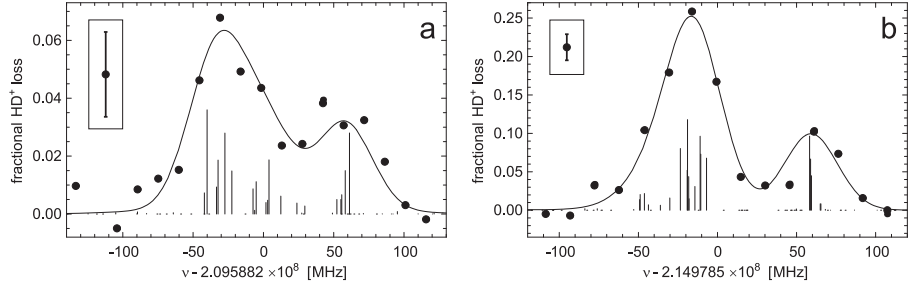}
\end{center}
    \caption{\label{22fig:results}
    Rovibrational transition spectra with partially resolved hyperfine
    splitting:
    (a) $(v\!=\!0,L\!=\!2)\to(v'\!=\!4,L'\!=\!1)$ at 1430 nm,
    (b) $(v\!=\!0,L\!=\!2)\to(v'\!=\!4,L'\!=\!3)$ at 1394 nm.
    The curves are fits to the data ($\bullet$), where the theoretical stick
    spectra were broadened by $\approx$40 MHz.
    The theoretical spectrum exhibits a large number of very weak transitions,
    due to weak mixing of pure coupled angular momentum states.
    The ordinate values are the molecular ion dissociation probability for a
    5 s irradiation of  0.65 W/cm$^2$ IR and 10 W/cm$^2$ UV light. The insets
    show typical error bars \cite{22Schiller06}.}
\end{figure}

\index{HDplus@HD$^+$ ions!photodissociation}
The loss of HD$^+$ ions not only depends on the REMPD process, but also on
transitions induced by blackbody radiation (BBR). We modeled the loss of
HD$^+$ by solving the rate equations for the populations of all $(v,L)$
levels interacting with the IR and UV lasers, as well as with the BBR
radiation at 300 K. The theoretically obtained excitation spectrum (see
Fig.~\ref{22fig:results} and text below) of the levels probed by the IR
laser is included, but for the remainder of the calculation hyperfine
structure, due to electron, nuclear and rotational spins, is ignored. The
rovibrational transition moments involved are taken from
\cite{22Colbourn1976}. The rate of dissociation by UV light is obtained
using cross sections from \cite{22Tadjedine1977}. For typical UV
intensities, dissociation rates of $10^2$--$10^3$ s$^{-1}$ are found. The
rate equation model reveals two different timescales at which the HD$^+$
number declines during a typical experiment. A first, fast ($<\!1$ s) decay
occurs when the IR laser selectively promotes HD$^+$ ions from a specific
$(v\!=\!0,L)$ level to a rotational level in $v'\!=\!4$, from which they
are efficiently photodissociated. This process rapidly dissociates those
$(v\!=\!0,L)$ HD$^+$ ions which are in the hyperfine states probed by the
IR laser. The remaining molecular ions (a significant fraction of the
total initial number) are dissociated significantly slower, essentially at
the rate at which the hyperfine levels of $(v\!=\!0,L)$ are repopulated by
BBR and spontaneous emission. For example, for the $(v\!=\!0,L\!=\!2) \to
(v'\!=\!4,L'\!=\!1)$ transition, and for typical intensities of 6 W/cm$^2$
for the IR and 10 W/cm$^2$ for the UV laser, the fast HD$^+$ decay takes
place at a rate $\sim$10 s$^{-1}$ (which is not resolved experimentally),
whereas the decay due to BBR--induced repopulation occurs at a rate of
$\sim$0.04 s$^{-1}$. The latter rate is fairly consistent with the
measured decay depicted in Fig.\ref{22fig:method}(b), but observed decay
rates depend strongly on which part of the hyperfine spectrum is
interrogated. This points at a shortcoming of the simple rate equation
model used here, and our observations can probably be explained precisely
only by a rate equation model which takes the full hyperfine structure of
all involved $(v,L)$ levels into account.

As an example, Fig.~\ref{22fig:method}(b) shows the time evolution of the
HD$^+$ secular excitation resonance while the HD$^+$ ions are excited on
the maximum of the rovibrational line
$(v\!=\!0,L\!=\!2)\to(v'\!=\!4,L'\!=\!1)$ at 1430.3883 nm. The decrease of
the HD$^+$ resonance in the secular excitation spectrum, induced by the
REMPD process, is accompanied by a decrease of the dark crystal core
containing the MHIs. The secular excitation spectrum
also shows an increase of the number of D$^+$ ions, which result from the
dissociation of excited HD$^+$ ions. These ions are sympathetically cooled
and remain in the crystal core. Fig.~\ref{22fig:method}(c) shows the
mixed-species ion crystal after all HD$^+$ was dissociated. The dark
crystal core has shrunk significantly, and the crystal now contains
$\approx$1100 Be$^+$ and $\approx$50 D$^+$ ions. Assuming equal
probability for photodissociation to D$^+$ and H$^+$, this number
indicates that most generated D$^+$ ions are sympathetically cooled and
trapped. Loss rates are obtained by exponential fitting to the maxima of
the HD$^+$ resonances in the secular excitation spectrum (solid line in
Fig.~\ref{22fig:method}(b)). In this way, a 0.01 s$^{-1}$ background loss
rate of HD$^+$ ions from the trap is obtained when both the IR and UV
lasers are turned off. This loss is due to chemical reactions between
HD$^+$ ions and background gases. The observed background loss rate is
fitted well by a single exponential decay, which rules out strong
nonlinear dependence of the Be$^+$ fluorescence during secular excitation
on the number of HD$^+$ ions.

The spectroscopic signal used to produce the spectra in
Fig.~\ref{22fig:results} is the molecular ion dissociation probability,
obtained as the relative change of the heights of the HD$^+$ secular
resonances in the Be$^+$ fluorescence before and after the REMPD
excitation (Fig.~\ref{22fig:method}(d)). For each transition, the HD$^+$
dissociation probability was measured as a function of the frequency of
the IR laser, in steps of 15 MHz. Each data point was obtained by
averaging over several individual measurements of the HD$^+$ dissociation
probability occurring over $\sim$5 s. Each data point requires a new
loading of HD$^+$ ions in the Be$^+$ crystal. For all measurements,
comparable HD$^+$ ion numbers were used, as deduced from the size of the
crystal core after loading. However, during each HD$^+$ loading cycle a
small fraction of the Be$^+$ is lost from the trap, due to chemical
reactions with neutral HD gas \cite{22RothPRA2006}. The same Be$^+$ ion
crystal can be used for up to 40 HD$^+$ loadings, sufficient for obtaining
the spectra in Fig.~\ref{22fig:results}. A typical spectrum is taken within
1-2 hours.

\index{Hydrogen molecular ions!hyperfine structure} \index{HDplus@HD$^+$
ions!effective spin Hamiltonian} Detailed measurements for two transitions
$(v\!=\!0,L\!=\!2)\to(v'\!=\!4,L'\!=\!1,3)$ are shown in
Figs.~\ref{22fig:results}(a,b). Both spectra reveal a partly resolved
hyperfine structure, which can be compared with the prediction from an
effective spin Hamiltonian, written as $H_{\rm eff} = b_1 {\bf I}_p \cdot
{\bf S} + c_1 I_{pz} S_z + b_2 {\bf I}_d \cdot {\bf S} + c_2 I_{dz} S_z +
\gamma {\bf S} \cdot {\bf J}$ \cite{22Ray1977,22Carrington1989}. Here, ${\bf
I}_p$, ${\bf I}_d$, and {\bf S} denote the spin of the proton, deuteron,
and electron, respectively; the subscript $z$ indicates the projection on
the internuclear axis. The hyperfine coefficients $b_1$, $b_2$, $c_1$,
$c_2$, and $\gamma$ have been recently calculated to high accuracy
\cite{22HD_hfs}, see Sec.~\ref{22subsec:hfs}. The hyperfine level energies and
eigenfunctions are found by diagonalization of the matrix representation
of $H_{\rm eff}$ in a suitable angular momentum coupling scheme. Terms
arising from the nuclear spin-rotation and deuteron quadrupole
interactions are neglected as they contribute $\ll$1 MHz to the hyperfine
level energies \cite{22HD_hfs}. The results of the diagonalization were
subsequently used to calculate line strengths (Eq.~\ref{22eq:intens}) of
the individual hyperfine components within a given rovibrational
transition, leading to "stick spectra", as shown in
Fig.~\ref{22fig:results}. Inhomogeneous broadening of the spectra may be
accounted for by convolving each line with a Gaussian lineshape of a given
width.

The broadened stick spectra are fitted to the experimental spectra using
the linewidth, the vertical scale and the frequency offset as fit
parameters \linebreak(Fig.~\ref{22fig:results}). The frequency offset
corresponds to the deperturbed ro-vib\-ra\-tio\-nal transition frequency,
which is thus determined to within the accuracy of the wavelength
calibration of the IR laser (40 MHz) and the fit uncertainty (3 MHz). The
measured deperturbed ro-vibrational transition frequency is in good
agreement with the {\em ab initio} results from \cite{22Moss1993}, see
Fig.~\ref{22fig:results}. The partly resolved hyperfine structure in the
measured spectra agrees well with the theoretical results obtained from
\cite{22Ray1977,22HD_hfs}. We find both theoretically and experimentally that
the hyperfine structure for other transitions in the P and R branches is
similar to that in Figs.~\ref{22fig:results}(a,b).

\index{HDplus@HD$^+$ ions!Doppler broadening}
\index{Micromotion of ions in an ion crystal}
We observe a typical line broadening of 40 MHz, which corresponds
to $k_B \times (0.2\,\hbox{\rm K})$ of energy in the axial motion.
The kinetic energy in the secular motion (as inferred from
molecular dynamics simulations) of the HD$^+$ ions can give rise
to broadening of about 10 MHz only \cite{22Blythe2005}. Saturation
broadening also does not play a significant role, as confirmed by
comparing spectra taken at different IR and UV intensities. Using
the polarization-dependent 313 nm fluorescence of the Be$^+$ ions
as a magnetic field probe, the magnetic field (which is along the
direction of propagation of the 313 nm laser beam) has been
adjusted and verified to be 50 mT and to vary by no more than 40
mT over the extent of the crystal, which implies Zeeman broadening
$<$1 MHz. This leaves Doppler broadening due to micromotion as the
most probable cause for the observed line broadening. This
micromotion could arise from phase shifts in the rf potentials
applied to the various electrodes, and from coupling between axial
(IR laser beam direction) and radial ion motion. For our trap, in
which the HD$^+$ ions are located at least 10 $\mu$m away from the
trap axis, the (radial) micromotion energy exceeds $k_B$($0.5\;$K).

\index{HDplus@HD$^+$ ions!Lamb-Dicke regime}
The results described are of significance in several respects.
They demonstrate, for the first time, the possibility of
high-resolution spectroscopy of small, trapped molecular ion
samples, sympathetically cooled well into the millikelvin range.
We have achieved a spectral resolution 10 times higher than with
any previous cold molecular ion method, and the same enhancement
was obtained for the excitation rate. The observed population
dynamics demonstrated the weakness of collisions. The methods used
for trapping, cooling and detection are quite general, and are
applicable to a host of other molecular ion species. This includes
other ions of astrophysical and cosmological interest such as
H$_3^+$ and its isotopomers, which have been trapped in our setup
\cite{22Blythe2005,22RothJPhysB2006}. Also, the spectral resolution
achieved here may be further improved: for instance, first-order
Doppler broadening may be circumvented by use of a tightly
confining trap which holds the ions in the Lamb-Dicke regime, or
by two-photon spectroscopy. Furthermore, the presence of the
atomic coolant ions offers an in situ tool to detect possible
perturbing fields.

\section{Conclusion and Outlook}

\index{Systematic shifts}
In summary, the development of high-accuracy laser spectroscopy of trapped
MHIs has made significant progress. On the theory side, the energies
have been calculated with a relative accuracy of the order of 1 ppb.
Detailed predictions of the line strengths of one- and two-photon
transitions have been given, which are important guides for the
experiments. Certain systematic shifts (dc and ac Stark shifts
\cite{22paris1,22paris2}) have also been calculated, but are not described
here. On the experimental side, several important techniques have been
demonstrated: cooling of MHIs to tens of mK, vibrational-state selective
photodissociation, one-photon vibrational spectroscopy with spectral
resolution at the level of $2\cdot10^{-7}$, rotational population
measurement, in-situ ion detection, tunable, high-power, continuous-wave
narrow-linewidth laser for two-photon spectroscopy. Based on the present
results, it is expected that the two-photon H$_2^+$ spectroscopy
experiment will ultimately allow a spectral resolution at the level of
$3\cdot10^{-10}$, while the one-photon 1.4\,$\mu$m HD$^+$ spectroscopy in
the current apparatus will be limited by Doppler broadening to several
parts in $10^8$. One-photon spectroscopy of HD$^+$ vibrational transitions
having longer wavelength or the use of a trap with stronger confinement
should allow reaching the Lamb-Dicke regime, with a strong increase in
spectral resolution.  As described above, two-photon spectroscopy is
another alternative.

For both ion species, the investigation of systematic shifts will become an
important task. It is expected that Zeeman shifts and Stark shifts can be
reduced or measured to a level below one part in $10^{10}$ in a cold ion
ensemble. This should enable comparisons of experimental and theoretical
transition frequencies at levels below 1 ppb, and, in the longer term, the
development of a novel approach to the measurement of mass ratios of
electron and hydrogen isotopes.

In the future, it may become attractive to use the method of
quantum-logic-enabled spectroscopy \cite{22Wineland2002,22Schmidt2005}. Some of
the experimental limitations (broad state population distribution, need for
destructive detection of molecular excitation, systematic effects)
encountered with the approaches described here could be substantially
alleviated.

\section*{Acknowledgements}

We thank P.~Blythe and H.~Daerr for their contributions and M.~Okhapkin,
A.~Nevsky, I.~Ernsting and A.~Wicht for discussions and assistance. This
work was supported by the German Science Foundation, the EC Network
HPRN-CT-2002-00290 "Cold Molecules". We also thank S.~Kilic, F.~Bielsa,
A.~Douillet, T.~Valenzuela and O.~Acef for their contributions, as well as
the LNE-SYRTE (Paris Observatory) and LPL (Universit\'e Paris 13)
laboratories for lending IR optical components and their absolute
frequency measurement setup. (J.K.) thanks the
Alexander-von-Humboldt-Foundation for support. (V.K.) acknowledges support
of RFBR, grant No.\ 05-02-16618.

%INDEX%%%%%%%%%%%%%%%%%%%%%%%%%%%%%%%%%%%%%%%%%%%%%%%%%%%%%%%%%%%%%%%
% Please check with the editor of your book whether he plans to
% include a "mutual" subject index - if so, please code your entries
% in the standard syntax. For your own purposes you may print your
% "personal" index by using the following commands:
%
%\clearpage
%\addcontentsline{toc}{section}{Index}
%\input{22idx.tex}
%\flushbottom
%\printindex
%%%%%%%%%%%%%%%%%%%%%%%%%%%%%%%%%%%%%%%%%%%%%%%%%%%%%%%%%%%%%%%%%%%%%

\end{document}